\documentclass[preprint,3p,number,review,sort&compress]{elsarticle}

\usepackage{psfrag}
\usepackage{amsmath,amssymb}
\usepackage{dsfont}
\usepackage{color}
\usepackage{nicefrac}
\usepackage{appendix}

\begin{document}


\title{Phase reconstruction from oscillatory data with iterated 
Hilbert transform embeddings - benefits and limitations}

\author[up]{Erik Gengel}
 \ead{egengel@uni-potsdam.de}
\author[up,unn,hse]{Arkady Pikovsky}%
\ead{pikovsky@uni-potsdam.de}
\address[up]{%
 Institute for Physics and Astronomy,  University of Potsdam, 
 Karl-Liebknecht Str. 24/25,
 14476 Potsdam, Germany 
}%
\address[unn]{Department of Control Theory,  Institute of Information Technologies,
Mathematics and Mechanics, Lobachevsky University Nizhny Novgorod, Russia}
\address[hse]{National Research University Higher School of Economics,  Nizhny Novgorod, Russia}
\begin{abstract}
In the data analysis of oscillatory systems, methods based on phase reconstruction are widely 
used to characterize phase-locking properties and inferring the phase dynamics. The main 
component in these studies is an extraction of the phase from a 
time series of an oscillating scalar observable. We discuss a practical procedure of phase 
reconstruction by virtue of a recently proposed method termed \textit{iterated Hilbert 
transform embeddings}. We exemplify the potential benefits and limitations of the approach by 
applying it to a generic observable of a forced Stuart-Landau oscillator. Although in many cases, 
unavoidable amplitude modulation of the observed signal does not allow for perfect phase 
reconstruction, in cases of strong stability of oscillations and a high frequency of the forcing, 
iterated Hilbert transform embeddings significantly improve the quality 
of the reconstructed phase. We also demonstrate 
that for significant amplitude modulation, iterated embeddings do not provide any improvement.
\end{abstract}

\maketitle

\section{Introduction}

The quantitative analysis of oscillatory phenomena by means of non-linear methods has become an indispensable tool in  widespread research areas such as physics \cite{Matheny_etal-19, blackbeard2014synchronisation,Nixon_etal-13},
chemistry \cite{ocampo2019weak,Omelchenko-Sebek-Kiss-18,blaha2011reconstruction},  engineering \cite{strogatz2005crowd}, life 
sciences \cite{Lowet_etal-16,Meij_etal-15,benitez2001use, topccu2018disentangling, Kralemann_etal-13,YELDESBAY201994}, 
geoscience \cite{battista2007application, zappala2020mapping}, 
and communication \cite{shahal2020synchronization}. 
Various procedures of analysis combine statistical and state-space methods or seek to reconstruct a full dynamical model from data \cite{Kantz-Schreiber-04}. 

For oscillatory systems, a commonly adopted approach
is to characterize oscillations through their phases. In the context of data analysis, a phase is reconstructed from an observed time series of the dynamics. The obtained phase is then used either for statistical characterization of synchronization, and of other interdependencies such as network connectivity \cite{Lowet_etal-16},
or for a data-driven reconstruction of the phase dynamics \cite{Rosenblum-Pikovsky-01,Penny_et_al-09,RevModPhys.89.045001}. This latter approach resides on the rigor phase reduction theory, 
which allows for a description of the dynamics of weakly perturbed or coupled limit-cycle oscillators 
in terms of their 
phases \cite{Kuramoto-84,pikovsky2001synchronization,Nakao-16, pietras2019network,gengel2020high}.

This paper focuses on the initial problem of extracting the phases from a scalar time series. 
We will not touch next possible steps (e.g., a characterization of the phase dynamics). 
The difficulties arising at the first stage of the analysis are manifold:
First, determination of the phase requires an embedding of the scalar signal into at least a two-dimensional state space. This means that an additional variable should be constructed from the available scalar signal.
Second,  an available scalar observable can have a pretty complex waveform, with several maxima and minima over the characteristic period. A good example is an electrocardiogram, with its several peaks (waves) during one cardiac 
cycle ~\cite{Kralemann_etal-13}. 
Therefore, one needs an approach that is universally applicable to complex oscillatory signals, and not only to those with a sine-like
waveform. 
Third, one has to distinguish between an arbitrary phase-like variable, a protophase, and the true phase, which potentially can always be defined for limit-cycle oscillators. 
The fourth and the most striking problem is that typical oscillatory signals possess modulations of
both the phase and the amplitude. To the best of our knowledge, there is no general observation-based method of
disentanglement of these modulations. This latter issue becomes especially critical in a widely-used approach to phase reconstruction
based on the Hilbert transform~\cite{feldman2011hilbert,king_2009}. Indeed,  the Hilbert transform is known
to ``mix'' the phase and the amplitude modulations of a signal \cite{bedrosian1962fm, bedrosian1962product, guevara2017neural, zappala2020mapping}. This, in particular, leads to the property 
that even for a purely phase modulated signal (i.e., if amplitude modulations are absent) 
the phase extraction is challenging. 
Recently, this latter problem, which we call phase demodulation problem, has been solved via an application of iterated Hilbert transform embeddings (IHTE) \cite{gengel2019phase}.
This method provides phase demodulation for complex waveforms and for broad modulation spectra (where previously one had to apply filtering or mode decomposition prior to demodulation~\cite{feldman2011hilbert, guevara2017neural, zappala2020mapping}).

This work aims to test the IHTE method on the observables from driven nonlinear limit-cycle oscillators.
Such general observables possess necessarily also amplitude modulation, and we will explore 
under which circumstances an application of the IHTE method is of merit. The 
following two steps will be carried out: First, we will extract a protophase from a generic, moderately complex scalar observable.  Such a protophase contains all available information about the true phase dynamics, but it does not coincide with the true phase. Thus, while it provides a proper estimation of the mean frequency
of oscillations, protophase-based calculations of the synchronization index and of the phase coupling functions are biased and depend on the method of protophase extraction.
In the second step, to obtain an invariant (i.e., independent on the details of the extraction)
representation of the phase dynamics, an additional protophase-to-phase transformation should be performed; this transformation is discussed in the last part of the paper.
\section{Nonlinear Limit-Cycle Oscillators and Phase Reduction} \label{sec: phase reduction}
\subsection{Phase Reduction: Theory}

Here we remind the theoretical framework behind the problem of phase dynamics reconstruction from data~\cite{Kuramoto-84,pikovsky2001synchronization,Nakao-16,pietras2019network}. 
An autonomous non-linear oscillatory system is described by an $N$-dimensional state variable $\mathbf{y}(t)$, which evolves in time according to the differential equation $\dot{\mathbf{y}} = \mathbf{f}(\mathbf{y})$. The system possesses an asymptotically stable limit cycle $\mathbf{y}_0(t)=\mathbf{y}_0(t+T)$ with period $T$ and frequency $\omega=2\pi/T$. In the basin of attraction of this cycle, the uniformly growing phase $\varphi(t) = \Phi[\mathbf{y}(t)]$ always exists and obeys the equation
\begin{equation}
\dot{\varphi} = \nabla_{\mathbf{y}} \Phi|_{\mathbf{y}(t)} \cdot \mathbf{f}(\mathbf{y}) = 
\omega = \frac{2\pi}{T}\;. \label{eq dot vp unperturbed}
\end{equation}
On the limit cycle, only the phase changes, so that the state of the 
system is fully determined by the phase: $\mathbf{y}_0(t)=\mathbf{y}_0(\varphi(t))$. 
Generally, in the basin of the cycle, one has to know the phase and the 
amplitudes (deviations from the limit cycle) to exactly determine the state of the system. 

 If the oscillator is perturbed by a small external force 
 $\varepsilon \mathbf{p}(\mathbf{y},t)$, 
 the dynamical equations are $\dot{\mathbf{y}} = \mathbf{f}(\mathbf{y}) + \varepsilon \mathbf{p}(\mathbf{y},t)$. 
 Here, the forcing term can be any bounded 
 regular, chaotic or stochastic function of time and state. 
 Substitution into equation for the phase yields
\begin{equation}
\dot{\varphi} = \omega + \varepsilon \nabla_{\mathbf{y}} \Phi|_{\mathbf{y}(t)} \cdot \mathbf{p}(\mathbf{y},t) \; . \label{eq: dot vp perturbed full}
\end{equation}
One can see that for a bounded driving force and small enough value of $\varepsilon$, the phase
grows monotonously $\dot{\varphi}>0$. For a non-smooth and non-bounded driving 
(e.g., for a Gaussian white noise forcing), 
an additional smoothing should be performed to ensure monotonicity of the phase.
The phase equation~\eqref{eq: dot vp perturbed full} is not a
closed equation for the phase $\varphi$, because the trajectory $\mathbf{y}(t)$ differs from the limit cycle solution $\mathbf{y}_0(\varphi)$.
However, for weak perturbations one can argue that deviations from the limit cycle are small 
\begin{equation}
\mathbf{y} = \mathbf{y}_0(\varphi)+\varepsilon\delta\mathbf{y}\;.
\label{eq:ampm}
\end{equation}
This allows for closing the equation for the phase~\eqref{eq: dot vp perturbed full} in the first order in $\varepsilon$, by substituting into \eqref{eq: dot vp perturbed full} the zero-order expression $\mathbf{y}\approx \mathbf{y}_0(\varphi)$:
\begin{equation}
\dot{\varphi} = \omega + \varepsilon \nabla_{\mathbf{y}} \Phi|_{\mathbf{y}_0(\varphi)} \cdot \mathbf{p}(\mathbf{y}_0(\varphi),t) = \omega + \varepsilon Q(\varphi,t) \;.
\label{eq dot vp perturbed}
\end{equation}
Formulae~\eqref{eq:ampm} and~\eqref{eq dot vp perturbed} describe the dynamics in the first order in $\varepsilon$. 

The equation of the phase dynamics \eqref{eq dot vp perturbed} accounts for the major component of the system response, corresponding to the neutrally stable flow direction along the limit cycle. Accordingly, deviations of the phase are not small (can be larger than $2\pi$), only the rate of their variations is small $\sim\varepsilon$. In contradistinction, the deviations of the amplitudes according to Eq.~\eqref{eq:ampm} are small. Therefore, to describe relevant phenomena like synchronization, one concentrates on the phase dynamics \eqref{eq dot vp perturbed} only.

\subsection{Phase Reduction: Reconstruction Problem} \label{sec: pr recons prob}
In the context of data analysis, the equations of a driven oscillatory system are usually 
unknown. Instead, one rather observes its dynamics from the outside. Then, the ultimate goal 
is to reconstruct the phase dynamics in the form of Eq.~\eqref{eq dot vp perturbed} from the observed time 
series as a low dimensional representation of the original dynamics. For this task, one typically has at 
hand a scalar observable, which is a function of the system state $X(\mathbf{y})$. For a perturbed dynamics 
close to the limit cycle \eqref{eq:ampm} this function can be expressed as
\begin{equation}
X(\mathbf{y}(t))= X(\mathbf{y}_0(\varphi)+\varepsilon\delta\mathbf{y})\approx
X(\mathbf{y}_0(\varphi))+
\varepsilon \left. \nabla_{\mathbf{y}} X\right|_{\mathbf{y}_0(\varphi)} \cdot \delta\mathbf{y}\;.
\label{eq:ampphm}
\end{equation}
One can see that generally the observed signal is a process with phase modulation 
(the first term on the r.h.s. of \eqref{eq:ampphm}) and with amplitude modulation (the second term on the r.h.s.). The problem is to determine the phase $\varphi(t)$ from the time series of $X(t)=X(\mathbf{y}(t))$. We address this problem in the following sections, using a forced Stuart-Landau oscillator as a basic example.

\section{Driven Stuart-Landau oscillator}
\label{sec:dslo}

\subsection{Stuart-Landau Oscillator}

The Stuart-Landau  oscillator (SLO) is a paradigmatic model for non-linear oscillations. We write its dynamics in dimensionless variables, where the amplitude of the limit cycle and its frequency are both one:  
\begin{equation}
\begin{aligned}
\dot u &= -v  +\mu u (1-u^2-v^2)+\alpha v(u^2+v^2-1) \;,\\
\dot v&=u+\mu v (1-u^2-v^2)-\alpha u (u^2 +v^2-1)+\varepsilon P(t)\;. 
\end{aligned}
\label{eq SL}
\end{equation}
Parameter $\mu$ defines the stability of the limit cycle, parameter $\alpha$ determines the inclination of isochrons at the limit cycle and is a measure of non-isochronicity (this parameter is also responsible for a non-trivial dependence of the phase on $(u,v)$, see below). In polar coordinates $u(t) = R(t)\cos(\phi(t)),\; v(t) =R(t)\sin(\phi(t))$, the SL system~\eqref{eq SL} reads 
\begin{eqnarray}
\dot{R} &=& \mu R(1-R^2) + \varepsilon P(t) \sin(\phi) \label{eq: SL radial R}\;, \\
 \dot{\phi} &=& 1  - \alpha(R^2-1)+\varepsilon R^{-1} P(t)\cos(\phi)\; . \label{eq: SL radial phi}
\end{eqnarray}
One can easily check that the uniformly rotating phase $\varphi$ of the SLO is
\begin{equation}
\varphi=\Phi(\phi,R)=\phi-\alpha\mu^{-1}\ln(R),\qquad \dot \varphi = 1 + \varepsilon \frac{\cos(\phi) - \alpha \mu^{-1} \sin(\phi)}{R}P(t)\;. \label{eq: varphi SL}
\end{equation} 
An equation for a small deviation $\Delta R=R-1$ of the amplitude from the limit cycle reads
\[
\frac{d}{dt}\Delta R=-2\mu \Delta R+\varepsilon P(t) \sin(\phi)\;.
\]
It follows that the characteristic values of the deviation are $|\Delta R|\sim \varepsilon/\mu$. Thus, the amplitude
modulation vanishes in the limit of strong stability of the limit cycle $\mu\to\infty$. In this limit 
the amplitude stays constant $R\approx 1$ and $\phi(t) \approx \varphi(t)$. We will use this in our exploration of the effect of the amplitude modulation on the quality of phase reconstruction.

\subsection{Scalar Observable}
As discussed above, to model observations of the dynamics, we need to define a ``generic'' scalar observable. We cannot expect this to be one of the dynamical variables of the underlying equations. Below we use a moderately complex observable:
\begin{equation}
X(u,v)=u^3+v+2uv\;.
\label{eq:obs}
\end{equation}
This observable is a smooth function of the state space variables $u(t)$ and $v(t)$, and it is complex in the sense that two maxima and minima exist on the basic period of oscillations (see Fig.~\ref{fig:am}).

\subsection{Forcing}
We choose a quasi-periodic forcing with three incommensurate frequencies (to avoid synchronization that would lead to a trivial phase dynamics):
\begin{equation}
P(t) =  \Big[ \cos(\Omega t)+\cos(\sqrt{3/5} \Omega t+\pi /4) +\cos(\sqrt{2/5} \Omega t+\pi /2))\Big] /3\;.
\label{eq:force}
\end{equation}
Parameter $\Omega$ allows us to test for low- and high-frequency forcing, in comparison with the basic frequency $\omega=1$ of the oscillator. In most numerical examples below we use $\varepsilon=0.3$.

\subsection{Levels of Amplitude Modulation}
\begin{figure}[!htbp!]
\centering 
\includegraphics[width=\columnwidth]{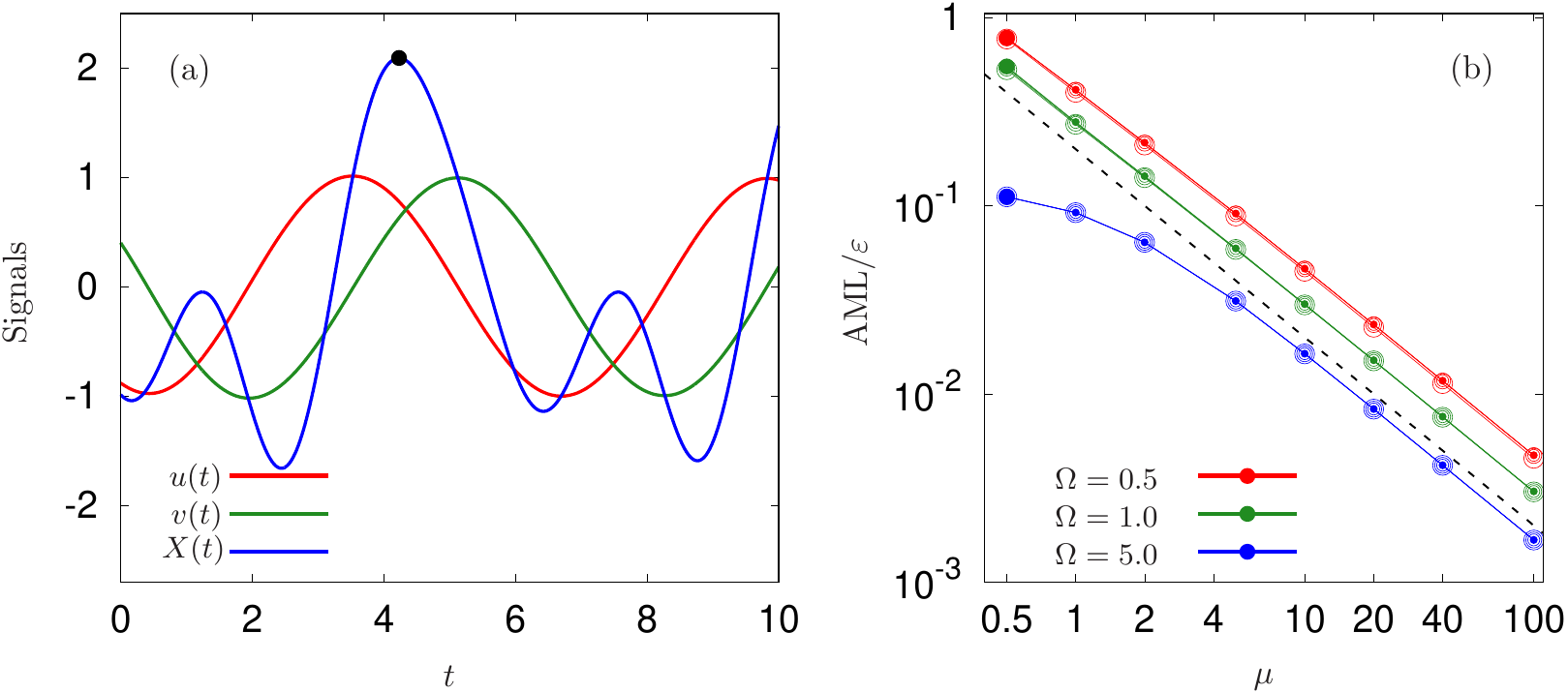}
\caption{Panel (a): The time series of variables $u(t),v(t)$ and of observable $X(t)$ (Eq.~\ref{eq:obs}) for the SLO with $\alpha=0.1$, $\mu=0.5$, $\Omega=0.5$ and $\varepsilon=0.1$. The bold black dot indicates a marker event that corresponds to a phase of $k 2\pi$, this marker event will be used below for the protophase definition. Because only a short piece of about one period is depicted, the modulation is not visible at this scale. Panel (b): Levels of the amplitude modulation, AML (cf.~Eq.~\eqref{eq:err1}), in the driven SLO Eqs.~\eqref{eq SL}, \eqref{eq:force} with observable from Eq.~\eqref{eq:obs}. The curves show AML normalized by the amplitude of the forcing $\varepsilon$, for four values of $\varepsilon=0.01,\;0.02,\;0.05,\;0.1$ (different dot sizes, they overlap nearly perfectly) and for three values of the forcing frequency as indicated in the panel. The dashed line shows the power law $0.2 \mu^{-1}$.}
\label{fig:am}
\end{figure}

For the SLO, an explicit relation between its phase $\varphi(t)$ and the state $\mathbf{y}(t)$ exists (Eq.~\eqref{eq: varphi SL}). Thus, we find the phase of the driven oscillations from the numerically obtained trajectory by virtue of Eq.~\eqref{eq: varphi SL}. In other situations, where the dynamics of the system is known, one can find the phase by a purely numeric procedure \cite{gengel2020high}. In this way we can check the relative importance of the two terms entering Eq.~\eqref{eq:ampphm}. 
It is instructive to rewrite \eqref{eq:ampphm} as
\begin{equation}
X(\varphi,R)\approx X(\varphi,1)+\varepsilon \Delta X(\varphi,R)
\label{eq:phampmod}
\end{equation}
The first term on the r.h.s., describing pure phase modulation, is $2\pi$-periodic in $\varphi$ while the second term should be aperiodic. Thus, to characterize the effect of the second term on $2\pi$-periodicity, we define the amplitude modulation level (AML) for the observable $X(\varphi,R)$ by calculating 
the normalized second moment of $X(\varphi+2\pi,R)-X(\varphi,R)$:
\begin{equation}
(\text{AML})^2=\frac{\int [X(\varphi+2\pi,R)-X(\varphi,R)]^2 d\varphi }
{\int [X(\varphi,R)-\langle X\rangle]^2 d\varphi }\;.
\label{eq:err1}
\end{equation}
Because the phase $\varphi$ grows nearly uniformly, the integrals in this expression can also be performed as time integrals. 
In Fig.~\ref{fig:am} we show this error normalized by the forcing strength $\varepsilon$, for different values $\mu$ and different frequencies of the forcing.

In case of the SLO, the curves for different $\varepsilon$ nearly perfectly overlap, showing that the AML is $\propto\varepsilon$. This level decreases with parameter $\mu$ as $\propto \mu^{-1}$. 
Thus, in the limit $\mu\to\infty$, the observable is purely phase modulated.  

\section{Protophases and the Phase Reconstruction Problem}

 According to the theory summarized in Section~\ref{sec: phase reduction}, the phase is defined as a $2\pi$-periodic variable which grows uniformly if the dynamics is unperturbed, i.e.~if amplitude perturbations vanish. For example, for the SLO \eqref{eq SL}, if variables $(u(t),v(t))$ on the limit cycle are observed,
 the definition $\phi=\text{arg}(u+iv)$ provides the true phase. If, however, the observed variables are slightly shifted,
 then the same definition $\phi'=\text{arg}(u+u_0+i(v+v_0))$ will provide a phase variable which is not growing uniformly in time, although it is $2\pi$-periodic. This observation motivated authors of Refs.~\cite{kralemann2007uncovering,kralemann2008phase} to introduce a notion of \textit{protophase} as a variable that parametrizes motion on a limit cycle, which is $2\pi$-periodic and monotonously -- but not uniformly --
 grows in time. Generally, a protophase depends on the details of its definition (in the example above on the values
 of the offset $(u_0,v_0)$, but as a rule also on the observables used, on the way one defines a coordinate along a closed
 curve, etc.).
Thus, it is convenient to introduce a family of protophases~\cite{kralemann2007uncovering,kralemann2008phase}, which can be obtained from the phase via a transformation
\begin{equation}
\psi=\Psi(\varphi),\qquad \text{where}\quad \Psi(\varphi)=\Psi(\varphi+2\pi) + 2\pi\;,\;\;\Psi' >0\;.
\label{eq:ptpg}
\end{equation}
For any admissible function $\Psi$, the variable $\psi$ will be an admissible protophase, fulfilling the 
condition of $2\pi$-periodicity. Generally, one can assume that a reconstruction method delivers some 
protophase $\psi$, not the true phase $\varphi$~\cite{kralemann2007uncovering,kralemann2008phase}.

In fact, a reconstruction of a protophase already provides a very rich knowledge on the dynamics, because the genuine phase is related to it via a one-to-one transformation of class \eqref{eq:ptpg}. For example, one could reconstruct dynamical equations of type \eqref{eq dot vp perturbed} in terms of a protophase. The disadvantage is that these dynamical equations are not observable-independent (because protophase-dependent), and cannot be compared  directly with theory, in contradistinction to the equations for the true phase.

Therefore, the phase dynamics reconstruction problem can be divided into two steps:
\begin{enumerate}
\item From a generic observation $X(t)$ of a dynamical system with a limit cycle, infer a protophase $\psi(t)$ that is a function of the
genuine phase $\varphi$.
\item Having obtained a protophase $\psi$, find a transformation to the genuine phase $\psi\to\varphi$.
\end{enumerate}
We will address problem 1 in Sections~\ref{sec:phdem} and \ref{sec:protf} below.  First, in Section~\ref{sec:phdem}
we will consider the case of a purely phase-modulated observable and present the method of Iterated Hilbert Transform Embeddings (IHTE), which solves the problem~\cite{gengel2019phase}. Then in Section \ref{sec:protf} we will explore, how well this method works for observables with an additional amplitude modulation. Finally, in Section~\ref{sec:p2p}, we will discuss the protophase-to-phase transformation problem~\cite{kralemann2007uncovering,kralemann2008phase}.
 
\section{Phase Demodulation via Iterated Hilbert Transform Embeddings}
\label{sec:phdem}

Here we consider the problem of phase reconstruction for a purely phase-modulated signal. Because, as discussed above, generally we have no chance to obtain the true phase, we from the beginning write this signal as a function of a protophase
\begin{equation}
X(t)=S(\psi(t)),\quad S(\psi)=S(\psi+2\pi)\;.
\label{eq:phmod}
\end{equation}
This signal corresponds to observations of the state of an oscillator, if on the r.h.s. of expressions \eqref{eq:ampphm} and \eqref{eq:phampmod} only the first term is present. The $2\pi$-periodic function $S$ will be called waveform. Of course, if another protophase is used, the waveform changes as well. The goal is to find one protophase and one waveform from this family.

\subsection{Case of two observables}
\label{sec:twoobs}

The solution of this problem is simple, if a second observable $Y(t)=S_y(\psi(t))$, also only purely phase-modulated, 
would be available (this second observable should be of course at least partially independent of the first one).
Then, the trajectory $(X(t),Y(t))$ on the plane $(X,Y)$ will be, according  to the Whitney' embedding theorem~\cite{Adachi-93}, a closed curve (although possibly with self-crossings)
and the protophase $\psi(t)$ can be chosen as some 
parametrization along this curve. For example, the curve length
\begin{equation}
L(t) = \int_0^t \sqrt{ \dot{X}^2+ \dot{Y}^2}dt 
\label{eq: length} 
\end{equation}
yields a monotonously growing function of time, even if the curve has many loops (i.e.,~if the waveform is complex). 
The protophase can then be defined as
\begin{equation}
\psi(t)=2\pi \frac{L(t)}{\mathcal{L}}\;,
\label{eq:lprtp}
\end{equation} 
where $\mathcal{L}$ is the total length of the closed loop and thus can be obtained from data. As has been discussed previously in \cite{gengel2019phase}, 
it stands in close correspondence to the average periodicity of the signal in terms of $L$. We stress here that for determining the protophase \eqref{eq:lprtp} no \textit{a priori}
knowledge on the properties of the system is needed.

\subsection{Hilbert transform embedding}
\label{sec:hte}

If only one scalar time series $X(t)$ is available, one tries to´ obtain the second scalar time series $Y(t)$ from the first one. 
The most widely used method for this task is the Hilbert transform (HT): $ Y(t) = \hat{H}[X(t)]$, where
\begin{equation}
\hat{H}[X(t)] = \frac{\text{p.v.}}{\pi}\int_{t^{(0)}}^{t^{(m)}} \frac{X(\tau)}{t-\tau} d \tau\; . \label{eq: ht}
\end{equation}
(Formally, the HT is defined on an infinite time interval, we write here finite limits $t^{(0)}$, $t^{(m)}$ because we will apply it to a finite time series). 

After calculating $Y(t)$, one performs an embedding on the plane $(X,Y)$. The main problem with this approach is that -- already for a purely phase-modulated 
signal -- the HT embedding is not a closed curve, but rather a band (see Fig.~\ref{fig:perfpdm} a,b). The reason for this is the well-known fact that the HT 
mixes modulations of amplitude and phase \cite{bedrosian1962fm, bedrosian1962product, guevara2017neural, zappala2020mapping}. Thus, any definition of a protophase based on the embedding $(X,\hat{H}[X])$ is not accurate.

\subsection{Iterated Hilbert transform embeddings}
\label{sec:ihte}

To resolve the problem, in Ref.~\cite{gengel2019phase} we demonstrated, theoretically and numerically, how an iteration of the HT embedding yields an almost perfect phase demodulation for purely phase modulated signals. In this procedure, remaining errors arise due to a discrete implementation of the HT and the finiteness of the time series. The main idea is to use the approximate phase, obtained from the embedding $(X,Y)$, as a new ``time'' variable, and perform a new HT in terms of this new time.

To describe the procedure algorithmically, it is convenient to introduce a protophase at each step of iterations $\theta_n$, 
and to treat time as the zero-order protophase $\theta_0=t$. Then, at each step $n$ one has a signal $X(\theta_n)$. The HT of this signal is calculated according to 
\begin{equation}
Y(\theta_n)=\hat{H}[X(\theta_n) ]= \frac{\text{p.v.}}{\pi}
\int_{\theta_n^{(0)}}^{\theta_n^{(m)}} \frac{X(\tilde{\theta}_n)}{\theta_n-\tilde{\theta}_n}\ d \tilde{\theta}_n\; . \label{eq:httheta}
\end{equation}
Here one should take into account that for different $n$ the functions $X(\theta_n)$ are different. For simplicity of notations, we keep this difference only by writing the corresponding index at the argument, the same holds for $Y(\theta_n)$. After $Y(\theta_n)$ is calculated, we perform an embedding in the plane $(X(\theta_n),Y(\theta_n))$ and calculate the (non-normalized) protophase at the next step as the length along the embedded trajectory, like in \eqref{eq: length}:
\begin{equation}
L_{n+1}(\theta_n) = \int_0^{\theta_n}\sqrt{(dX/d\theta_{n}')^2+ (dY/d\theta_{n}')^2}d\theta_{n}' \label{eq:ltheta} \; .
\end{equation}

In principle, for further iterations it is not necessary to normalize $L_{n+1}(\theta_n)$, because the 
outcome of the integration Eq.~\eqref{eq:httheta} is independent on the rescaling of the argument. 
It is, however, convenient for the sake of comparison and of presentation, to have at each step a protophase 
normalized to intervals of $2\pi$. We employ an approach based on  a spline interpolation. We define 
for each period one marker event (in the context of stochastic processes,
a slightly different, statistical concept of marker events is used \cite{callenbach2002oscillatory,Rice-44}). 
Practically, we take as this event the largest local maximum of the waveform on a 
period (as depicted in Fig.~\ref{fig:am} (a)). Subsequent markers correspond to 
phases $0,2\pi, 4\pi,\ldots,k 2\pi,\ldots$. If these events correspond to the 
values 
of the quantities calculated in Eq.~\eqref{eq:ltheta} 
$L_{n+1}^{(0)},  L_{n+1}^{(1)},L_{n+1}^{(2)}, \ldots,L_{n+1}^{(k)},\ldots$, 
then we define $\theta_{n+1}$ as a smooth function of $L_{n+1}$, having values $k 2\pi$ at 
values of the argument $L_{n+1}^{(k)}$. 
We denote this spline normalization of the length-defined protophase as
\begin{equation}
\theta_{n+1}(t)=\text{SPL}(L_{n+1}(t))\;.
\label{eq:spl}
\end{equation} 
We accomplish the construction with standard cubic splines \cite{press1992numerical}. One can easily see, that if the embedding in the plane $(X(\theta_n),Y(\theta_n))$ is a well defined curve, then this definition of the protophase $\theta_{n+1}(t)$ coincides with Eq.~\eqref{eq:lprtp} (because in this case the function \eqref{eq:spl} is a straight line). Notice, that in this procedure one calculates the values of the protophase $\theta_n(t)$ at the same instants, at which the initial time series was defined. In this way, applying steps Eqs.~\eqref{eq:httheta} - \eqref{eq:spl} iteratively for $n=0,1,2\ldots$, one obtains a series of the protophases $\theta_{n+1}(t)$ as functions of time. As has been demonstrated in Ref.~\cite{gengel2019phase}, this procedure converges to an almost perfectly $2\pi$-periodic protophase. 

Before applying this method to observations, of the driven SLO, we illustrate it with a purely phase-modulated signal like in Eq.~\eqref{eq:phmod}.

\subsection{Example - A purely phase-modulated signal}\label{sec: purely phase modulated}

Here we demonstrate properties of the IHTE using a purely phase modulated
signal. We create it from the dynamics of the forced SLO (Section \ref{sec:dslo}).
We use the basic observable \eqref{eq:obs}, but insert there purely phase-dependent variables
\[
\overline{u}(t)=\cos(\varphi(t)),\quad \overline{v}=\sin(\varphi(t)),\quad  \overline{X}=\overline{u}^3+
\overline{v}+2\overline{u}\overline{v}\;.
\]

\begin{figure}
\centering
\includegraphics[width=0.48\columnwidth]{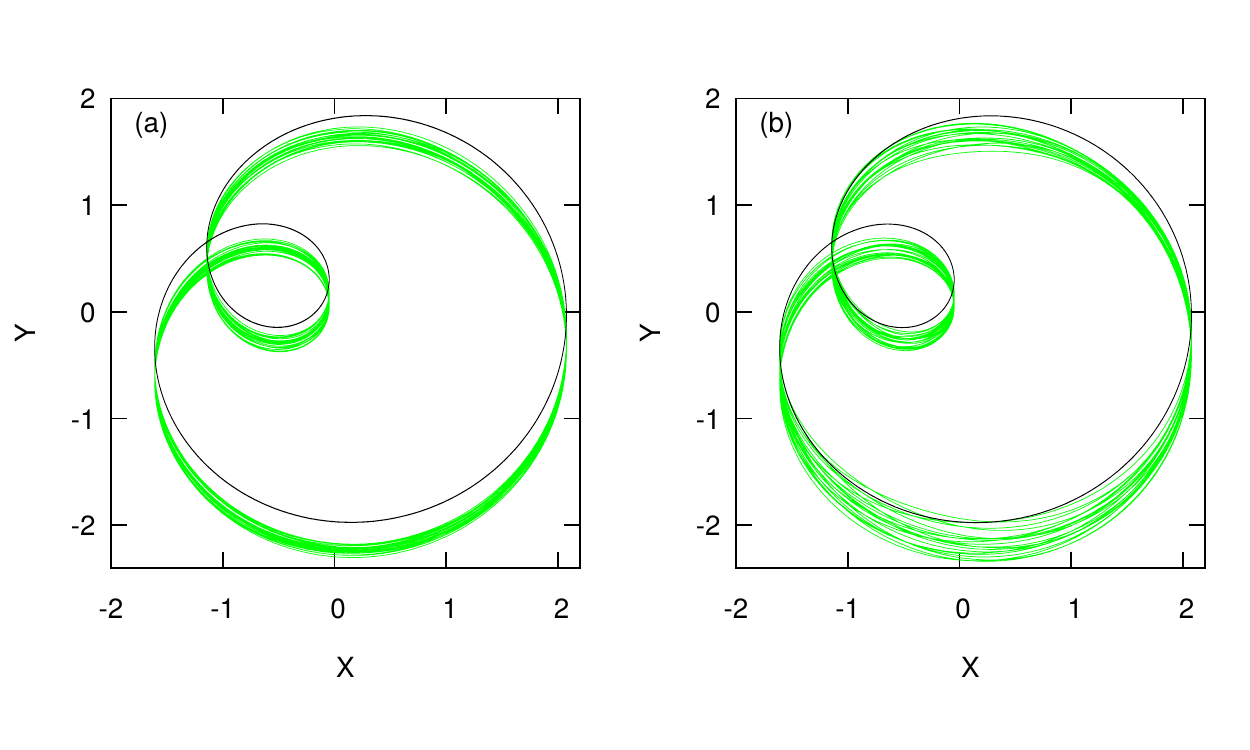}\hfill
\includegraphics[width=0.48\columnwidth]{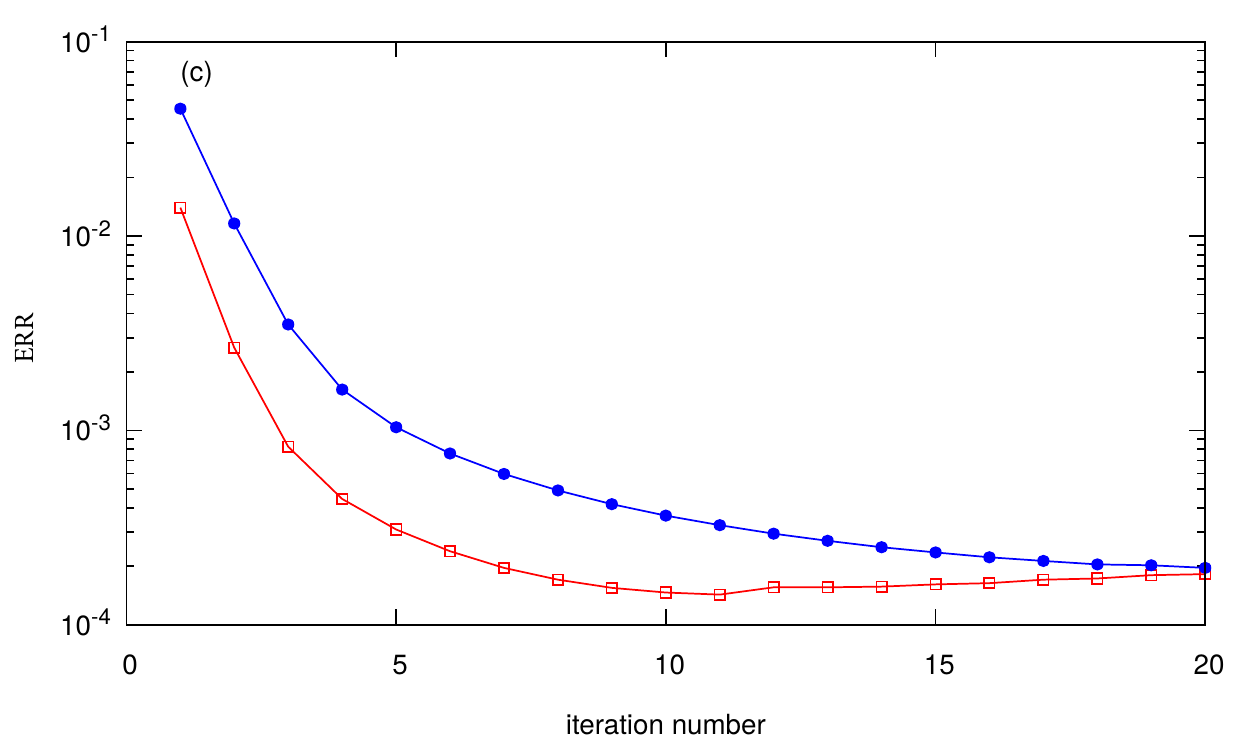}
\caption{ Panels (a) and (b): illustration of embeddings in plane $(\overline{X}(\theta_n(t)),\overline{Y}(\theta_n(t))$ for the first iteration $(n=1)$ (green line) and for 
the final iteration $n=20$ (black line). Panel (a): $\Omega=1$; panel (b): $\Omega=4$. One can see that at the first iteration a certain level of amplitude modulation appears 
(one observes a band of a finite width). At $n=20$ the line is nearly perfect, indicating for a very low level of residual amplitude modulation. Panel (c): Characterization 
of quality of the phase demodulation by periodicity error ERR  \eqref{eq:err2}. Filled circles: $\Omega=4$, open squares: $\Omega=1$.}
\label{fig:perfpdm}
\end{figure}

We characterize the success of the phase demodulation procedure in two ways, at each step of the iterative procedure where the approximate 
protophase $\theta_n(t)$ is determined. The first method is just a visualization of the embedding $(\overline{X}(\theta_n(t)),\overline{Y}(\theta_n(t)))$. 
For a perfect demodulation, this embedding produces a closed loop describing the waveform of the signal (cf. discussion in Section \ref{sec:twoobs}). 
In a more quantitative approach, the success of demodulation is determined by the periodicity error 
\begin{equation}
(\text{ERR}_n)^2=\frac{\int [X(\theta_n+2\pi)-X(\theta_n)]^2  d\theta_n}
{\int [X(\theta_n)-\langle X\rangle]^2 d\theta_n}\;.
\label{eq:err2}
\end{equation}
This expression is similar to the AML Eq.~\eqref{eq:err1}. We stress here,
that this expression is based on the observations only and does not require \textit{a priori} knowledge of the system.
The difference is that it now monitors the $2\pi$-periodicity of the waveform $X(\theta_n)$ which is an indicator for 
the remaining spurious amplitude modulation of the reconstructed waveform in step $n$. If $2\pi$-periodicity is achieved, 
then ERR$_n$ is zero. 

We illustrate the results for the driven SLO \eqref{eq SL},\eqref{eq:force} 
with $\varepsilon=0.3$ and two values of the basic frequency of the forcing $\Omega \in \left\{1,4\right\}$ 
in Fig.~\ref{fig:perfpdm}. Compared to the basic period of oscillations, these forcing frequencies 
give rise to a slow (case $\Omega=1$) and a fast (case $\Omega=4$)
forcing, respectively. The obtained results confirm findings of Ref.~\cite{gengel2019phase}, which can 
be summarized as follows: (i) For a slow (i.e.~$\Omega\lesssim 1$) and weak  (i.e.~$\varepsilon \ll \omega$)
modulation, 
already the first iteration (i.e. the traditional HT embedding) provides the 
highest periodicity of the protophase (e.g., in  Fig.~\ref{fig:perfpdm}(c) ERR$_1$ in the 
first iteration for $\Omega=1$ is approximately 4 times smaller than for $\Omega=4$, with the 
same amplitude of the forcing); (ii) for all other cases, i.e.~for a slow (i.e.~$\Omega\lesssim 1$) but strong (i.e., $\varepsilon \lesssim \omega$) modulation, 
as well as for a fast modulation (i.e.~$\Omega>1$), several iterations are needed 
to achieve a good demodulation. The final level of the error is limited by errors 
in numerical performance of the HT due to discreteness and finiteness of the signal.

\section{Protophase reconstruction from the full signal}
\label{sec:protf}

\begin{figure}
\centering
\includegraphics[width=0.6\columnwidth]{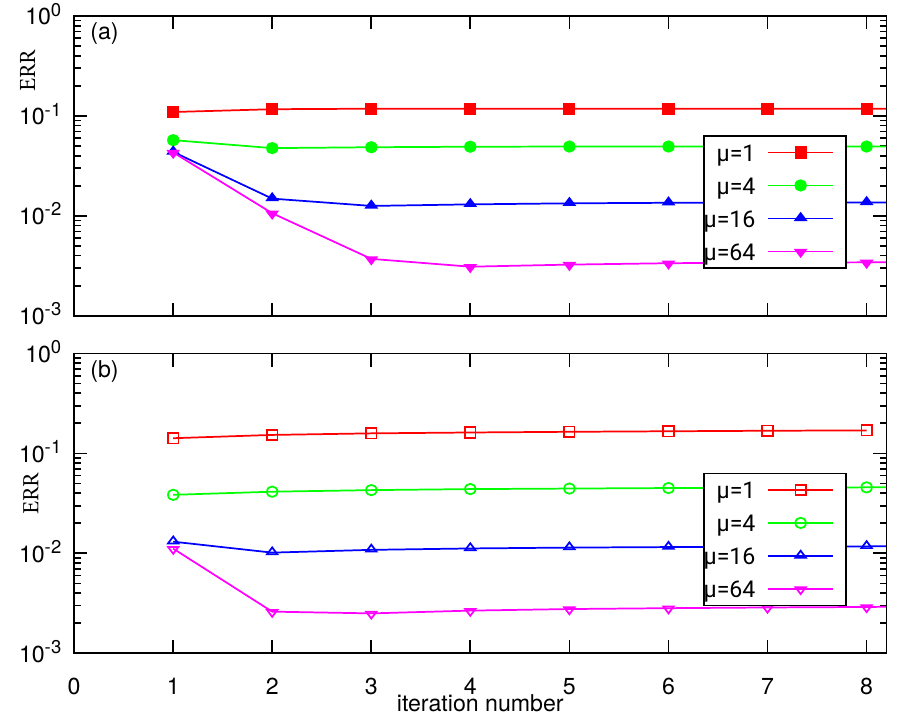}
\caption{ Periodicity error ERR$_n$ for the phase reconstruction by virtue of the IHTEs. Panel (a): $\Omega=4$, panel (b) $\Omega=1$.}
\label{fig:prq1}
\end{figure}

\begin{figure}[!htbp!]
\centering 
\includegraphics[width=0.7\columnwidth]{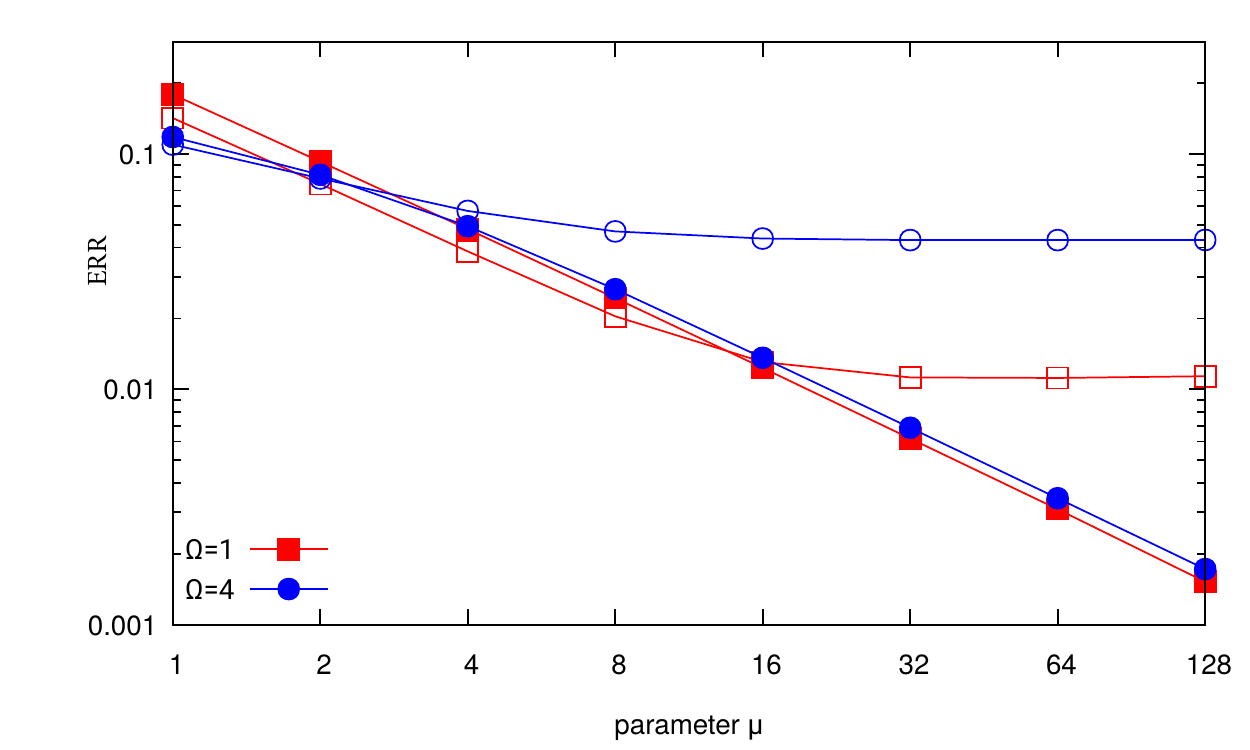}
\caption{Periodicity error ERR$_n$ in the protophase reconstruction in dependence on parameter $\mu$: ERR$_1$ at iteration step $n=1$ is shown with open markers, and error Err$_{20}$ at iteration step $n=20$ is shown with corresponding filled markers.}
\label{fig:prq2}
\end{figure}

Here we report on the application of the IHTE procedure to the time series of the observable $X(t)$ \eqref{eq:obs} from the forced SLO. As the observable 
contains both a phase and an amplitude modulation, we cannot expect that the resulting modulation error \eqref{eq:err2} will be small; in fact it is 
bounded below by the amplitude modulation level. The question we would like to address is whether iterations improve the quality of the protophase 
reconstruction. We show the results in Figures \ref{fig:prq1}, \ref{fig:prq2}. Figure \ref{fig:prq1} shows the error according to \eqref{eq:err2} 
versus the iteration number for different values of the driving frequency $\Omega$ and of the parameter of  stability of the limit cycle $\mu$. 
One can see that for a slow forcing (panel (b), $\Omega=1$), a weak improvement is achieved for large $\mu=64$ only, i.e., for a weak amplitude 
modulation. In all other cases, already the first iteration provides a reconstruction that cannot be further improved. On the other hand, for 
higher frequency of the forcing (panel (a)), and for large enough values of stability parameter $\mu\in \left \{16,64 \right \}$, iterations 
do indeed improve the quality of the reconstruction - by one order of magnitude for $\mu=64$. This general conclusion is supported also 
by Fig.~\ref{fig:prq2}, which shows, in dependence on $\mu$, the periodicity error ERR$_n$ after the 1st and the last ($n=20$) iteration steps. 
From this data one can see how significant the improvement is due to IHTE for extremely large stability of the limit cycle ($\mu=128$). An improvement is already present for $\mu>4$ for $\Omega=4$, and for $\mu>16$ for $\Omega=1$.

\section{Protophase-to-Phase Transformation}\label{sec:p2p}

After having obtained a protophase $\psi(t) = \theta_{n}(t)$ from IHTE as described in Section~\ref{sec:ihte} above, the true phase needs to be estimated. We assume that any good protophase is a function of the true phase $\psi = \Psi(\varphi)$. Thus, the problem can be reformulated in terms of finding the inverse protophase-to-phase (PTP) transformation $\varphi=\Psi^{-1}(\psi)$.

\subsection{Two variants of PTP transformation}

In Ref.~\cite{kralemann2007uncovering,kralemann2008phase}, a method  
to reconstruct the protophase-to-phase transformation (PTP) $\Psi^{-1}(\psi)$ has been suggested. It is based on the assumption that the phase probability distribution density $\rho(\varphi)$ should be uniform: $\rho(\varphi)=(2\pi)^{-1}$. This condition indeed selects the true phase for an undriven oscillator: because $\dot\varphi=\omega$, the density of $\varphi$ is uniform.  Next, one assumes that under small forcing the distribution is only slightly perturbed such that uniformity should be valid at least up to corrections $\sim\varepsilon$~\cite{kralemann2007uncovering,kralemann2008phase}.
 
Practically, implementation of the PTP according to the condition of uniformity is based on the estimation of the density of the protophase $\rho(\psi)$~\cite{kralemann2007uncovering,kralemann2008phase}. It is convenient to represent this density via $K$ Fourier modes, the amplitudes $F_k$, $k=1,\ldots,K$ of which are calculated directly from the time series of the protophase $\psi(t)$:
\begin{equation}
F_k = \frac{1}{2\pi} \int_0^{2\pi} \rho(\tilde{\psi}) \exp(-ik\tilde{\psi}) d\tilde{\psi} = \frac{1}{t^{(m)}-t^{(0)}} \int_{t^{(0)}}^{t^{(m)}} \exp(-ik \psi(t)) dt \; .
\label{eq:ptp1}
\end{equation}
After this, the PTP transformation is accomplished according to the expression~\cite{kralemann2007uncovering,kralemann2008phase}
\begin{equation}
\varphi = \Psi^{-1}(\psi) + \xi^{db}(\psi)=\int_0^{\psi(t)} \rho(\tilde{\psi}) d \tilde{\psi} + \xi^{db}(\psi) = \psi + \sum^{K}_{k\neq 0} 
\frac{F_k}{ik}\Big(\exp(ik\psi) -1 \Big) + \xi^{db}(\psi) \;.
\label{eq:ptp2} 
\end{equation}
The residual term of this density-based PTP is denoted by $\xi^{db}(\psi)$. 

The PTP transformation \eqref{eq:ptp1},\eqref{eq:ptp2} has the advantage that it is purely data-based; 
thus, on the one hand, no additional information is needed. On the other hand, it is based on a condition 
(uniform density of the true phase), that is fulfilled only approximately.  To evaluate 
the quality of the data-driven PTP, we additionally estimate the coefficients of the mapping $\Psi^{-1}(\psi)$ 
from the least square fit to $\varphi$, according
to the following expression:
\begin{equation}
\Big \langle \left[\varphi(t)-\psi(t) - \sum^{K}_{k=0} A_k \cos(k\psi(t)) + B_k \sin(k\psi(t))\right]^2\Big \rangle_t \rightarrow \text{min}\; .
\label{eq: direct fitting PTP}
\end{equation}
(We used the \verb|C++| library \verb|EIGEN| to perform this task.) The residual term of fitting here is 
denoted as $\xi^{fit}(\psi)$. We would like to stress that this approach makes use 
of $\varphi(t)$ and is thus based on theoretical information, not available in an experiment. However, 
it can be deemed as the optimal PTP because it is not bound to the condition of uniformity for $\rho(\psi)$ 
discussed before.

\subsection{Accuracy of phase reconstruction}

\begin{figure}[!htbp!]
\centering 
\includegraphics[width=0.32\columnwidth]{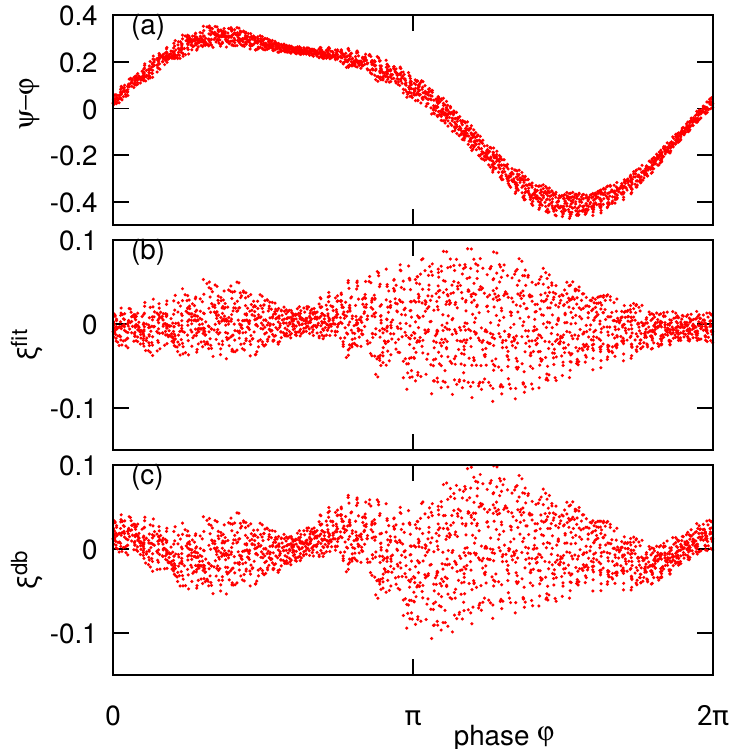}\hfill
\includegraphics[width=0.32\columnwidth]{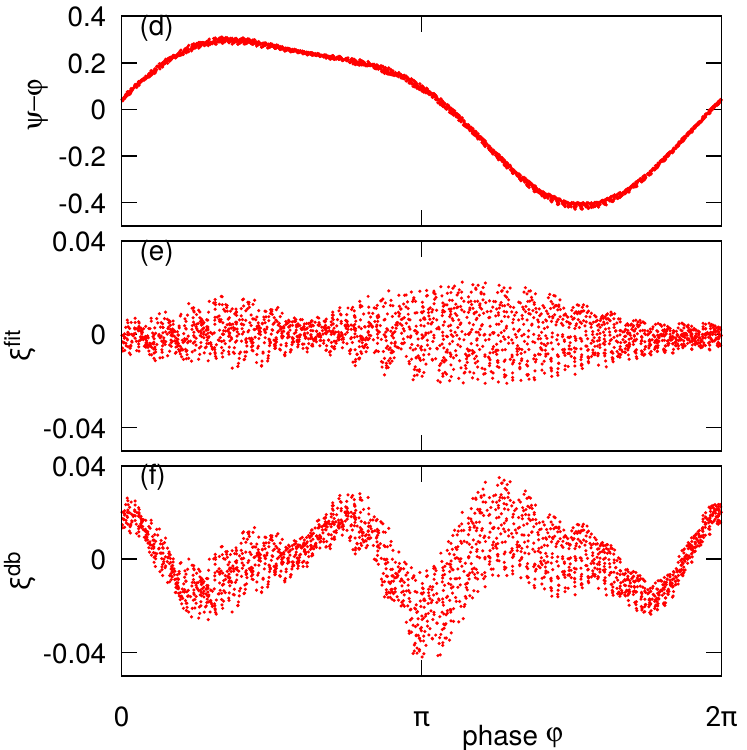}\hfill
\includegraphics[width=0.32\columnwidth]{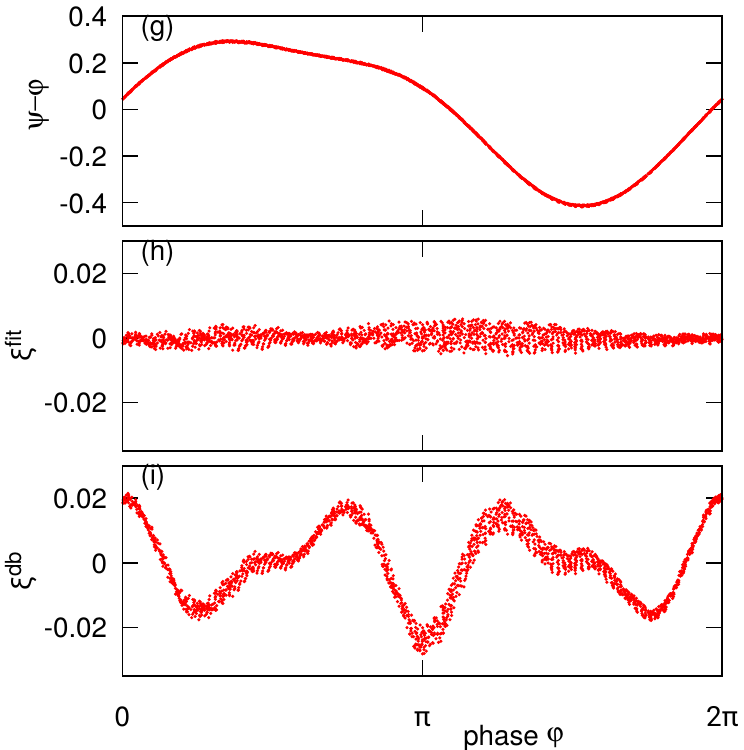}
\caption{Panels (a,d,g): Differences $\psi-\varphi$ obtained in the IHTE procedure vs 
the true phase $\varphi$. Here, $\psi=\theta_{20}$ after 20 iterations. Panels (b,e,h): 
Residua $\xi^{fit}(\psi(\varphi))$ obtained by applying to the protophase $\psi$ the fitting procedure 
\eqref{eq: direct fitting PTP}. Panels (c,f,i): Residua $\xi^{db}(\psi(\varphi))$ obtained by applying 
to the protophase $\psi$ the PTP procedure \eqref{eq:ptp2}. Parameter of the forcing is $\Omega=4$, 
parameters of the SLO: in panels (a,b,c) $\mu=4$, in panels (d,e,f) $\mu=16$, and in panels (g,h,i) 
$\mu=64$. Notice that the panels showing the residua have different vertical scales for different 
values of $\mu$. In all panels, many periods of the signal are overlapped.
}
\label{fig:fit}
\end{figure}

Here in Fig.~\ref{fig:fit} we present exemplary residua $\xi^{fit,db}(\psi(\varphi))$ of phase reconstructions after PTP from the direct fitting Eq.~\eqref{eq: direct fitting PTP} and 
from the density-based approach Eq.~\eqref{eq:ptp2} (both with $K=30$) for signals $X(t)$ at different frequencies of the forcing, $\Omega$, and of the oscillator stability $\mu$.  
In accordance with the results of Section \ref{sec:protf} above, only for forcing frequencies larger than the base frequency, and for large values of parameter $\mu$, i.e. 
for strong stability of the limit cycle, the iterations of the IHTE allow for an essential improvement. Moreover, for the case of low-frequency forcing, the two variants of 
the PTP transformation give very similar results. We attribute this to the fact that the uncertainty in the protophase reconstruction is significantly larger than the violation 
of the uniformity condition for the density of the true phase.

For the high-frequency forcing, the results of the two variants of the PTP transformation are slightly different. It is instructive to compare panels (h) (direct fit according 
to \eqref{eq: direct fitting PTP}) and (i) (density-based transformation according to \eqref{eq:ptp2}). For the direct fit, the residual error does not depend on average on $\varphi$, 
thus it is intrinsic and cannot be removed by an additional transformation. In contradistinction, the curve of the density-based fit shows pronounced oscillations and could be 
potentially improved. This residual waveform is due to non-exactness of the condition of uniform density of the true phase.  

\begin{figure}
\centering
\includegraphics[width=0.8\columnwidth]{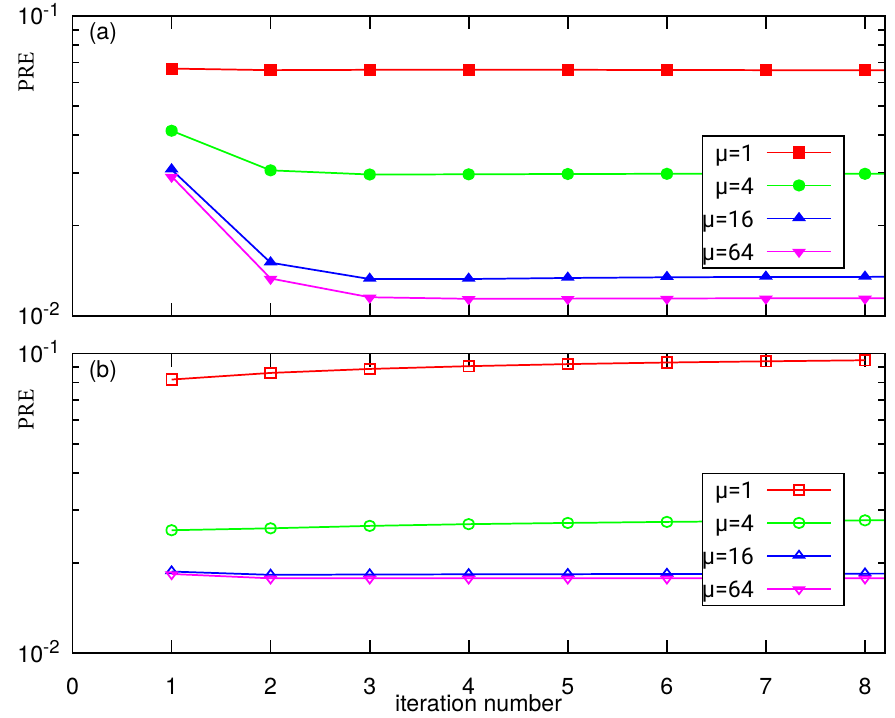}
\caption{Reconstruction error \eqref{eq:pre} for phase estimation by virtue of the IHTEs and subsequent
PTP transformation according to \eqref{eq:pre}. Panel (a): $\Omega=4$, panel (b) $\Omega=1$. The values of the SLO parameter $\mu$ are depicted in the panels.}
\label{fig:pre1}
\end{figure}

Finally, to judge the accuracy of the combined procedure of IHTE and PTP transformation, we define the phase reconstruction error PRE as the variance of the difference $\Psi^{-1}(\theta_n) - \varphi$:
\begin{equation}
\text{PRE}_n^2=\langle (\Psi^{-1}(\theta_n)-\varphi -\langle \Psi^{-1}(\theta_n)-\varphi\rangle)^2 \rangle,  
\label{eq:pre}
\end{equation}
and plot it vs the iteration number in Fig.~\ref{fig:pre1}. For a small forcing frequency ($\Omega=1$), practically no improvement via iterations is achieved. 
One can see here that the overall accuracy grows significantly with parameter $\mu$, i.e. with the increasing stability of the limit cycle. Panel (a) shows 
that for a higher forcing frequency ($\Omega=4$), the reconstruction error reduces by a factor of two to three within the first five iterations.

\begin{figure}[!htbp!]
\centering 
\includegraphics[width=\columnwidth]{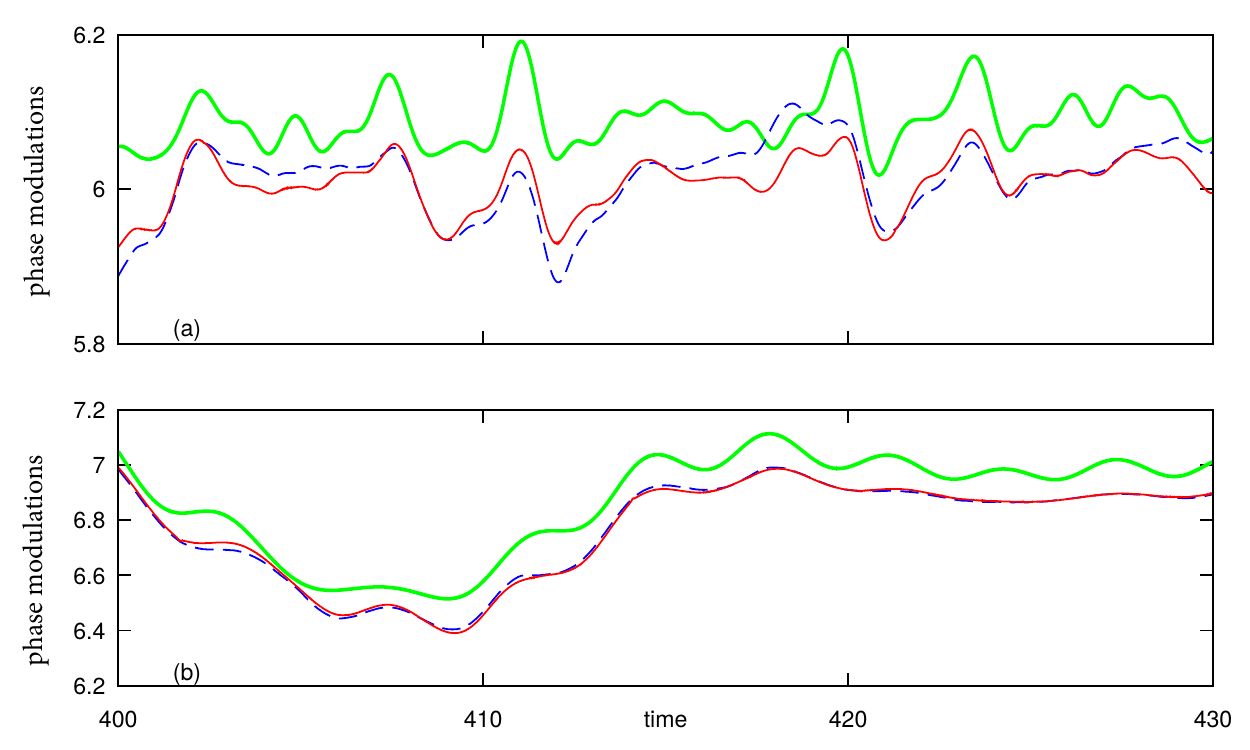}
\caption{Comparison of the true phase (green line) with phases obtained after the protophase reconstruction at the 1st iteration of the IHTE procedure 
(blue dashed line) and at iteration 20 (red line). To the protophases also the density-based PTP transformation according to expressions \eqref{eq:ptp1} 
has been applied. We removed the linear trend $\sim t$ from all phases and we shifted the true phase vertically for better comparison. Panel (a): $\Omega=4$, panel (b): $\Omega=1$.
}
\label{fig:modulations compare x xbar}
\end{figure}

As a visual example for the accuracy of reconstruction, 
Fig.~\ref{fig:modulations compare x xbar} depicts the true phase
together with reconstructions at the 1st and at the 20th iterations. 

\section{Conclusion}

The goal of this study was to explore how phase reconstruction from scalar observables 
 of oscillatory dynamical systems can be improved by the application of iterated Hilbert transform embeddings. 
 We considered the simplest model of stable self-oscillations, the Stuart-Landau oscillator, under quasiperiodic external forcing. We explored a range of the 
frequencies of the forcing and of the limit cycle stability parameter. The main conclusion is the following: In the case of purely phase-modulated signals, 
the true phase can be nearly perfectly reconstructed. In the more realistic situation where modulations of the amplitude and of the phase are present, the 
 improvement from IHTEs is moderate if any.

Because the amplitude modulation is heavily suppressed in the case of highly stable limit cycles, 
we have observed that for large values of $\mu$ IHTEs indeed enhance the quality of phase reconstruction. 
However, for small $\mu$ there is no merit in performing iterations. The reason for this is that the Hilbert transform (HT) ``mixes'' phase and amplitude modulations, 
and iterations beneficial in the purely phase-modulated case are spoiled by the intrinsic amplitude modulation of the signal. Practically, we could recommend 
performing at least one additional iteration of the HT embedding, and if it does not reduce the error, use just the phase from the first HT. Otherwise, one can perform several iterations until 
ERR$_n$ stops to decrease.
 Albeit there exists no strict correspondence between the periodicity error 
 ERR$_n$ Eq.~\eqref{eq:err2} and the reconstruction accuracy PRE$_n$ Eq.~\eqref{eq:pre}, 
 we suggest that the empiric criterion for ERR$_n$, put forward in this study, should 
 be used to test protophases for their quality ($2\pi$-periodicity).

 Here, we would like to stress several aspects of our approach 
 which are either not widespread in the literature or should be taken into account 
 when developing further the IHTE procedure:

\begin{enumerate}
\item We used a scalar observable of moderate complexity, which is a smooth function of the system variables. 
The observed signal thus does not have a simple sine-like form but has several maxima and minima on the basic period. Correspondingly, an embedding using HT leads to a ``band'' that has several loops. In such a situation one cannot use the argument of the analytic signal $X+i \hat{H}[ X]$ for the phase estimation. Instead, we used the length of the curve as an estimation of the protophase, following Refs.~\cite{Kralemann_etal-13,gengel2019phase}. 
\item It is important to distinguish a protophase, $\psi(t)$, which is the final result of successful phase demodulation (Section \ref{sec:phdem}), and an estimate of the true phase $\varphi(t)$. The true phase is just one out of a family of possible protophases, and typically does not arise through the demodulation procedure described above. While all the protophases are equivalent (up to an invertible transformation), 
knowing the true phase is of enormous merit, because this allows one to obtain from the data a description as 
close as possible to the theory. In this work, we used a density-based transformation from a length-defined protophase to the phase~\cite{kralemann2008phase} and demonstrated that it gives rise to small deviations from the ground truth. The remaining deviations arise due to the basic assumption behind this approach, namely that the true phase has a uniform distribution, which is valid only approximately. Due to this, we used different measures of accuracy to characterize the quality ($2\pi$-periodicity) of the protophase reconstruction (ERR$_n$), and the quality of the protophase-to-phase transformation (PRE$_n$) throughout iterations.
\item  We would like to point out that three levels of error estimation have been outlined 
in this paper. 
First, one can compare the width of the embedding curve in the first and the final step of iterations. This approach might be useful if just a rough idea is needed on how good demodulation is. Second, a quantitative and data-driven measure for the periodicity of a protophase is defined by the quantity ERR (Eq.~\eqref{eq:err2}). This measure does not assume any \textit{a priori} knowledge of the system's properties.
In contradistinction, the absolute reconstruction error PRE (Eq.~\eqref{eq:pre}) 
is measured directly for the true phase  $\varphi$. It is thus not applicable in studies of empirical data but can be calculated and compared to ERR if the dynamical equations are known.
\item  We would like to stress that iterated embeddings in the  IHTE are different from the embeddings used for a phase space representation of a high-dimensional attractor~\cite{sauer1991embedology}. 
Because we study (modulated) periodic oscillations, in the ideal case of pure phase modulation, the underlying object
is a one-dimensional manifold, which, according to Whitney's embedding theorem \cite{Adachi-93}, can be embedded in a two-dimensional plane. A Hilbert transform allows for a two-dimensional representation, but due to a mixture
of phase and amplitude modulations, it does not provide a good embedding at the first step. This drawback is cured
by adopting the iteration procedure according to Ref.~\cite{gengel2019phase}.
\end{enumerate}

As discussed above, resolving the amplitude modulation problem remains the main challenge in constructing an accurate phase reconstruction from data. However, it appears promising to attack this problem for situations where there is sufficient theoretical understanding of the amplitude dynamics in coupled oscillators. In particular, the theory of higher-order interactions of coupled Stuart-Landau oscillators~\cite{gengel2020high} includes explicit expressions for the amplitude variations in low orders in the coupling parameter; this information could be used to develop and validate methods of coping with the amplitude modulation. 

Moreover, we have shown that, despite a significant increase in the protophase periodicity due to IHTE, the pre-assumptions of data-driven PTP represent a potential obstacle for an accurate phase reconstruction. These results suggest improving the PTP procedure to non-uniform target phase densities and/or incorporating phase-amplitude coupling into the phase reconstruction. To the best of our knowledge, both approaches are unsolved problems. Particularly, the highly desirable data-driven reconstruction of a phase-amplitude coupling from geometric information of embedding curves is an open problem. Some work has been carried out given full 
observations of the limit cycle dynamics \cite{cestnikinfering2020}.

In this paper, we assumed that the available time series is an observable
of a perturbed oscillating dynamical system. Therefore we did not touch the issue of preprocessing the data
if this assumption is not fulfilled. In the case where the time series is a mixture of 
components from different oscillators, one could try to separate these components using signal decomposition methods (like empirical mode decomposition~\cite{battista2007application}, intrinsic mode functions~\cite{Stallone_etal-20},
Hilbert-Huang transform~\cite{hilberthuang}). However, it is not clear how such a preprocessing
affects the properties of the phase and the amplitude modulations that are crucial for phase extraction.
Quite popular are also methods of time-frequency analysis, based on the wavelet transform; they
aim at finding an instantaneous frequency of a signal~\cite{le2001comparison, Daubechies_etal-11}. These methods
typically provide a frequency averaged over several characteristic periods. Potentially,
one could reconstruct a protophase by integration of the instantaneous frequency. Comparing the wavelet-based methods and the Hilbert transform-based technique described above remains a subject for future
studies.

Finally, we would like to stress that the phase dynamics concepts have been extended for chaotic oscillators and systems with strong noise  \cite{schwabedal2012optimal, schwabedal2013phase}. However,
in these cases, the amplitude modulation of the observed signals is very strong, and we do not expect the IHTEs to improve the phase reconstruction task.

\section*{Acknowledgements}
E. G. thanks the Friedrich-Ebert Stiftung and the Potsdam Graduate School for financial support.  
A. P. was supported by Russian Science Foundation (Grant No. 17-12-01534). Numerical experiments in 
Sec. 5,6 were supported by the Laboratory
of Dynamical Systems and Applications NRU HSE of the Russian
Ministry of Science and Higher Education (Grant No. 075-15-2019-
1931).
We thank Michael Rosenblum, 
Michael Feldman, and Mathias Holschneider 
for their advices and fruitful discussions.


\end{document}